\documentstyle[preprint,prd,aps,12pt,epsf]{revtex}
\begin{document}
\draft
\preprint{
\vbox{
\halign{&##\hfil\cr
         & DESY-98-xxx \cr
         & hep-ph/9812204\cr
         & October 1998 \cr
         & \cr
         & \cr
}}}
\vskip 0.5cm
\title{The Supersymmetric QCD Radiative Corrections to Top Quark
Semileptonic Decays} 
\author{Cong-Feng Qiao\footnote{Alexander von Humboldt Fellow. e-mail:
qiaocf@mail.desy.de}}
\address{CCAST-World Laboratory, P.O. Box 8730, Beijing 100080, China.\\
II Institut f\"ur  Theoretische Physik, Universit\"at Hamburg, D-22761
Hamburg, Germany.\footnote{Corresponding author address; Tel:
+49-40-89982232(o); Fax: +49-40-89982267(o)}}
\author{Shou-Hua Zhu}
\address{CCAST-World Laboratory, P.O. Box 8730, Beijing 100080, China.\\
Institute of Theoretical Physics, Academia Sinica, P. O. Box 2735,
Beijing 100080, China.}
\maketitle
\vskip 5mm
\begin{abstract}
The one-loop supersymmetric QCD corrections to the top quark semileptonic
decays $t\rightarrow b ~\bar{l} ~{\nu_l}$ are considered. The corrections
are found to reduce the decay width. In the still acceptable parameter
space the corrections to the percentage change in top quark semileptonic
decay width can reach the level of -0.7\%.

\vspace{4mm}
\end{abstract}
~\\
Keyword(s): SUSY QCD, RADIATIVE CORRECTIONS, TOP DECAY
\vfill \eject
\section{Introduction}
The discovery of the top quark in 1995 \cite{r1,r2,r3} marked the triumph
of the Standard Model(SM) of particle physics. Since the discovery, 
more interest on top quark physics research is stimulated in both
experimental and theoretical respects. Because of its large mass
which is close to the electroweak scale, the top quark physics research 
may play an important role in the study of the electroweak symmetry
breaking and therefore of the origin of the fermion masses. Through the
study on it people also hope to find clues of possible "new physics". 

Within the framework of SM the dominant top quark decay channel
is $t \rightarrow b W$, which has a time scale $\tau_W$ much shorter than
the time scale of the non-perturbative QCD effect. This character of top
quark makes it behaves almost like a free particle, which is helpful in 
the precise study on it. With the operation of CERN Large Hadron
Collider (LHC) in the future there will be enormous top quark events
obtained in every year and that provides it practically possible to
make the more detailed study on many properties of top quark.

Because of the importance of top quark decay stated above, a lot
theoretical efforts have been made in the studies of radiative corrections
to its decays. In the SM, the QCD and electroweak corrections to the $t
\rightarrow b W$ decay can be found in \cite{r4,r5}. Recently, the order
${\cal O}(\alpha_s^2)$ QCD radiative corrections on some specific
situations are presented in literatures \cite{r8,r9,r10}. Beyond the SM,
based on the two Higgs doublet model (2HDM) the radiative corrections to
$t \rightarrow b W$ are calculated in \cite{r11}. The supersymmetric 
(SUSY) contributions of both QCD and electroweak are given in \cite{r12}.

The top quark semileptonic decays is also one of the major and important
decay channels of the top quark, because there is a great portion of W
decay probability will proceed leptonic pattern. Before the top quark was
detected, it was long believed that the top mass maybe less than that of W
boson's. So, the $t\rightarrow b ~\bar{l} ~{\nu_l}$ processes \cite{r16}
were taken into consideration even earlier than $t \rightarrow b W$
process. And also the QCD corrections to it was completed \cite{r16a,r17},
which was found coincides with results in Ref.\cite{r4} after taking the
$W$ on its mass shell. In this paper we present a calculation on SUSY
QCD radiative corrections to the top quark semileptonic decays, which we
find is still not appeared in literatures.

\section{Formalism}

The relevant Feynman diagrams for the one-loop virtual gluino
corrections to the top quark semileptonic decay process 
$t(p) \rightarrow b(k_1)\bar l(k_2) v_l(k_3)$ are shown in Fig. 1. 
The tree-level amplitude (Fig.1(a)) can be written down straightforwardly
using the standard Feynman rules,
\begin{eqnarray}
M_{0} (t \rightarrow b ~\bar{l} ~{\nu_l})={2 \pi \alpha \over 
x_W(k^2-M_W^2+i \Gamma_W M_W)} [\bar u_b(k_1) \gamma_\mu P_L u_t(p)] [\bar
u_{\nu_l}(k_3)\gamma^\mu P_L v_{\bar l}(k_2)].
\end{eqnarray}
Here, $k=k_2 + k_3$ is the momentum transfer carried by the W boson, $P_L$
is the left-handed helicity projection operators, and the $x_W$ is the 
square of the sine of the Weinberg angle, $x_W={\rm sin}^2\theta_W$.

In one-loop calculations, we will use dimensional regularization to
control all the ultraviolet divergences in the virtual loop corrections
and we will adopt the on-mass-shell renormalization scheme \cite{r13,r14}.  
It is noted that dimensional reduction approach is widely used in the
calculations of the radiative corrections in the minimal supersymmetric
standard model (MSSM) as it automatically preserves the supersymmetry (at
least to one-loop order), while dimensional regularization does not. But
in the calculations of SUSY QCD corrections to the process concerned, the
dimensional reduction scheme is the same with the conventional dimensional
regularization one. The one-loop amplitude of the discussed process can be
written as
\begin{eqnarray}
\delta M=M^{con}+M^{unr},
\end{eqnarray}
where $M^{con}$ and $M^{unr}$ represent the counterterm and the 
un-renormalized amplitude, respectively. And,
\begin{eqnarray}
M^{con} &=& {1\over 2} (Z_{Lt}+Z_{Lb}) M_0,
\end{eqnarray}
where
\begin{eqnarray}
Z_{Lt} &=& {\alpha_s \over 4 \pi m_t^2} \{ \cos^2 \theta 
[(m_{\tilde{t}_2}^2-m_{\tilde{g}}^2) (B_0(2)-B_0(4))
-m_t^2 B_0(3) \nonumber \\
& &
+(m_t^2(-m_{\tilde{g}}^2+m_{\tilde{t}_2}^2-m_t^2) (
\dot B_0(3)+ \dot B_0(4))] + 4 m_{\tilde{g}} m_t^3 \cos\theta \sin\theta 
(\dot B_0(3)+ \dot B_0(4)) \nonumber \\
& &
+\sin^2\theta [(m_{\tilde{t}_1}^2-m_{\tilde{g}}^2) (B_0(1)-B_0(3))
-m_t^2 B_0(4) \nonumber \\
& &
+(m_t^2(-m_{\tilde{g}}^2+m_{\tilde{t}_1}^2-m_t^2) ( \dot B_0(3)+ \dot
B_0(4))]\}
\end{eqnarray}
and
\begin{eqnarray}
Z_{Lb} &=&Z_{Lt}[\tilde{t}\rightarrow \tilde{b},\theta \rightarrow 0] 
\end{eqnarray}
are the wave-function renormalization constants for top and bottom quark
respectively. In above and following equations the $B_0$, $C_0$, $C_i$,
and $C_{i,j}$ functions result from the evaluation of certain loop
integrals, expressions for which can be found in \cite{r13,add}. In 
equation (4), $\dot B_0(p^2,m_a^2,m_b^2)=\partial B_0(p^2,m_a^2,
m_b^2)/\partial p^2$, and $(1)-(4)$ are abbreviations of variables in
$B_0$ function, which represent $(0, m_{\tilde{g}}^2, m_{\tilde{t}_1}^2)$,
$(0, m_{\tilde{g}}^2, m_{\tilde{t}_2}^2)$, $(m_t^2, m_{\tilde{g}}^2,
m_{\tilde{t}_1}^2)$ and $(m_t^2, m_{\tilde{g}}^2, m_{\tilde{t}_2}^2)$,
respectively. Throughout our calculations, we will omit the bottom scalar
left- and right-handed mixing, but consider the top scalar mixing, because
of the small bottom quark mass compared with the top quark mass, the
definition of the mixing angle of top scalar $\theta$ can be found in next
section.

\begin{eqnarray}
M^{unr}&=& {\alpha_s \alpha \over 
x_W(k^2-M_W^2+i \Gamma_W M_W)}
\bar u_b(k_1) (f_1 \gamma_\mu P_L 
+f_2 k_{1\mu} P_R+ f_3 k_{1\mu} P_L+
f_4 \rlap/k k_{1\mu} P_L + f_5 p_{\mu} P_R
\nonumber \\
&+& 
f_6 p_{\mu} P_L+ f_7 \rlap/k p_{\mu} P_L + f_8 k_{\mu}P_R+ f_9 k{\mu}P_L
+ f_{10} \rlap/k k_{\mu} P_L) u_t(p) \bar u_{v_l}(k_3)\gamma_\mu P_L
v_{\bar l}(k_2),
\end{eqnarray}
where
\begin{eqnarray}
f_1 &=&2 (\cos^2\theta C_{00}(1)+\sin^2\theta C_{00}(2)), 
\end{eqnarray}
\begin{eqnarray}
f_2 &=&\sin\theta \cos\theta (C_{0}(2)- C_{0}(1)+2 C_1(2)-2 C_1(1)), 
\end{eqnarray}
\begin{eqnarray}
f_3 &=& m_b (\cos^2\theta (C_{0}(1)+3 C_1(1)+2 C_{11}(1)) 
+\sin^2\theta (C_{0}(2)+3 C_{1}(2)+2 C_{11}(2))), 
\end{eqnarray}
\begin{eqnarray}
f_4 &=& -\cos^2\theta (C_{2}(1)+2 C_{12}(1)) 
-\sin^2\theta (C_{2}(2)+2 C_{12}(2)), 
\end{eqnarray}
\begin{eqnarray}
f_5 &=&\sin\theta \cos\theta m_{\tilde{g}}
(C_{0}(1)- C_{0}(2)), 
\end{eqnarray}
\begin{eqnarray}
f_6 &=& -\cos^2\theta (C_{0}(1)+ C_{1}(1)) 
-\sin^2\theta (C_{0}(2)+ C_{1}(2)), 
\end{eqnarray}
\begin{eqnarray}
f_7 &=& \cos^2\theta C_{2}(1) 
+\sin^2\theta C_{2}(2), 
\end{eqnarray}
\begin{eqnarray}
f_8 &=& \sin(2 \theta) m_{\tilde{g}}(C_{2}(1) 
-C_{2}(2)), 
\end{eqnarray}
\begin{eqnarray}
f_9 &=& -2 ( \cos^2\theta (C_{2}(1)+C_{12}(1)) 
+\sin^2\theta (C_{2}(2)+C_{12}(2))), 
\end{eqnarray}
\begin{eqnarray}
f_{10} &=& 2 ( \cos^2\theta C_{22}(1) 
+\sin^2\theta C_{22}(2)). 
\end{eqnarray}
Here, (1) and (2) in the parentheses of above equations are the variables
of $C$ function which represent $(m_b^2, m_t^2, k^2, m_{\tilde{b}_1}^2
m_{\tilde{g}}^2, m_{\tilde{t}_1}^2)$ and $(m_b^2, m_t^2, k^2, 
m_{\tilde{b}_1}^2  m_{\tilde{g}}^2, 
m_{\tilde{t}_2}^2)$ for simplicity, respectively.

The corresponding amplitude squared of the process $t \rightarrow b 
~\bar l ~v_l$ can then be expressed as
\begin{eqnarray}
\overline{{\sum}}\left| M\right|^2=\overline{{\sum}}\left| M_{0}\right|^2
+2 Re (\overline{{\sum}}\delta M M_0^\dagger),
\end{eqnarray}

\section{Results and Discussion}

In our numerical calculations, we choose $\alpha_s=0.118$, $\alpha=1/128$,
$M_Z=91.187$ GeV, $M_W=80.33$ GeV, $\Gamma_W=2.1$ GeV, $m_t=176.0$ GeV,
and $m_b=4.5$ GeV, which are commonly used in literatures \cite{r15}.

In the MSSM the squark mass eigenstates $\tilde{q}_1$ and $\tilde{q}_2$  
are related to the current eigenstates $\tilde{q}_L$ and $\tilde{q}_R$ by   
\begin{eqnarray}   
\label{eq1}   
\left(\begin{array}{c}   
\tilde{q}_1 \\ \tilde{q}_2\end{array}\right)= R^{\tilde{q}} \left 
(\begin{array}{c} \tilde{q}_L \\ \tilde{q}_R\end{array}\right)
\end{eqnarray}
with
\begin{eqnarray}
R^{\tilde{q}}=\left(\begin{array}{cc} \cos\theta & \sin\theta\\
-\sin\theta & \cos\theta    
\end{array}\right).   
\end{eqnarray}   
The mixing angle $\theta$ and the masses $m_{\tilde{q}_{1,2}}$
of squarks can be obtained by diagonalizing the following mass matrices
\cite{r18}   
\begin{eqnarray}   
\label{eq2}   
M^2_{\tilde{q}}=\left(\begin{array}{cc} M_{LL}^2 & m_q M_{LR}\\   
m_q M_{RL} & M_{RR}^2   
\end{array} \right), 
\end{eqnarray}
where
\begin{eqnarray}
M_{LL}^2=m_{\tilde{Q}}^2+m_q^2+m_{z}^2\cos 2\beta 
(I_q^{3L}-e_q x_W),    
\end{eqnarray}   
\begin{eqnarray}
M_{RR}^2= m_{\tilde{U},\tilde{D}}^2 +m_q^2+m_{z}^2 e_q x_W \cos 2\beta,
\end{eqnarray}   
\begin{eqnarray}
\label{eq23}   
M_{LR}= M_{RL}=\left\{\begin{array}{ll} A_t-\mu \cot \beta &
(\tilde{q}= \tilde{t}),\\ A_b-\mu \tan \beta & (\tilde{q}= \tilde{b}).   
\end{array}   
\right.   
\end{eqnarray}   
In Eq.(\ref{eq2})-(\ref{eq23}), $ m_{\tilde{Q}}^2$, $m_{\tilde{U},
\tilde{D}}^2$ are soft SUSY breaking mass terms of the left- and
right-handed squark, respectively; $\mu$ is the coefficient of the $H_1
H_2$ mixing term in the superpotential; $A_t$ and $A_b$ are the
coefficients of the dimension-three trilinear soft SUSY-breaking term;
$I_q^{3L}, e_q$ are the weak isospin and electric charge of the squark
$\tilde{q}$. From Eqs.(\ref{eq1}) and (\ref{eq2}) $m_{\tilde{t}_{1,2}}$
and $\theta$ can be derived out  
\begin{eqnarray}   
m^2_{\tilde{t}_{1,2}}&=&{1\over 2}\left[ M^2_{LL} + M^2_{RR}\mp \sqrt{   
(M^2_{LL}-M^2_{RR})^2+4 m^2_t M^2_{LR}}\right],   
\end{eqnarray}   
\begin{eqnarray}   
\tan\theta&=&{m^2_{\tilde{t}_1}-M^2_{LL} \over m_t M_{LR}}.   
\end{eqnarray}   

The results of percentage change in the top quark semileptonic decay
width, $\delta \Gamma_{SL}/\Gamma_{SL}$, after considering the SUSY QCD
corrections are presented in Figs. 2-4. In numerical calculations we also
assume $\mu=-100$ GeV and $m_{\tilde{U}}=m_{\tilde{D}}= m_{\tilde{Q}}=A_t
=m_S$ (the so-called global SUSY). From the Tevatron dileptons and jet
+$\not\!\!\!{{E}_T}$ analysis under certain assumptions and parameter
inputs, people get that $m_{\tilde g}> 180$ GeV \cite{add1} and $m_{\tilde
q} > 90$ GeV (especially for the stop lower limit) presently \cite{add2}.

In Fig. 2 we show the relative corrections of one-loop SUSY QCD to the
$t\rightarrow b ~\bar{l} ~{\nu_l}$ decay widths as the function of $m_S$,
while keeping the $m_{\tilde g}$ and tan$\beta$ fixed. Taking $m_{\tilde
g} = 200$ GeV and tan$\beta=2$ we find that the total relative correction
for $l = e,~ \mu,~ \tau$ reaches -0.6\% from -0.33\% as the $m_S$ goes
from 100 GeV to 300 GeV. In Fig. 3 we take the $m_S$ and tan$\beta$ fixed,
$m_S=100$ and tan$\beta=2$, and show the relation of relative correction
with the $m_{\tilde g}$. It can be seen that the relative correction grows
up till -0.9\% with the decline of $m_{\tilde g}$ down to 150 GeV. In
Fig.4 we show the dependence of relative correction on tan$\beta$ when
take $m_S= 100$ GeV and $m_{\tilde g}=200$ GeV. The correction varies from
-0.6\% to -0.41\% with tan$\beta$ going from 2 to 50, and it is interesting 
to note that the curve gradually goes to be flat with the growth of
tan$\beta$, which indicates that the correction is not closely sensitive
to the value of tan$\beta$ as it larger than 10.

In conclusion, we have calculated the SUSY QCD corrections to the
semileptonic top quark decays. It is found that the corrections may reach
the level of about one percent with a favorable parameter choice.  
Because of the fact that the $t\rightarrow b ~\bar{l} ~{\nu_l}$ process
has a large branching ratio in top quark decays, with the run of LHC in
the future and then the large amount of top quark events obtained, we hope
the precise measurement of top quark semileptonic decay may give indirect
hints in the existence of SUSY. However, it should be noted that though
results in Fig.2-4 are independent of the top quark inclusive decay width,
without a calculation of the inclusive decay width based on the current
experimental allowed parameter choice information on the semileptonic
decay width alone can not be used to reach a conclusion on SUSY.

\vskip 1.2cm
\centerline{\bf \large Acknowledements}
This work was supported in part by the Natioinal Natural Science
Foundation of China, the Postdoctoral Foundation of China and K. C. Wong
Education Foundation, Hong Kong.

\begin{figure}
\epsfxsize=15 cm
\centerline{\epsffile{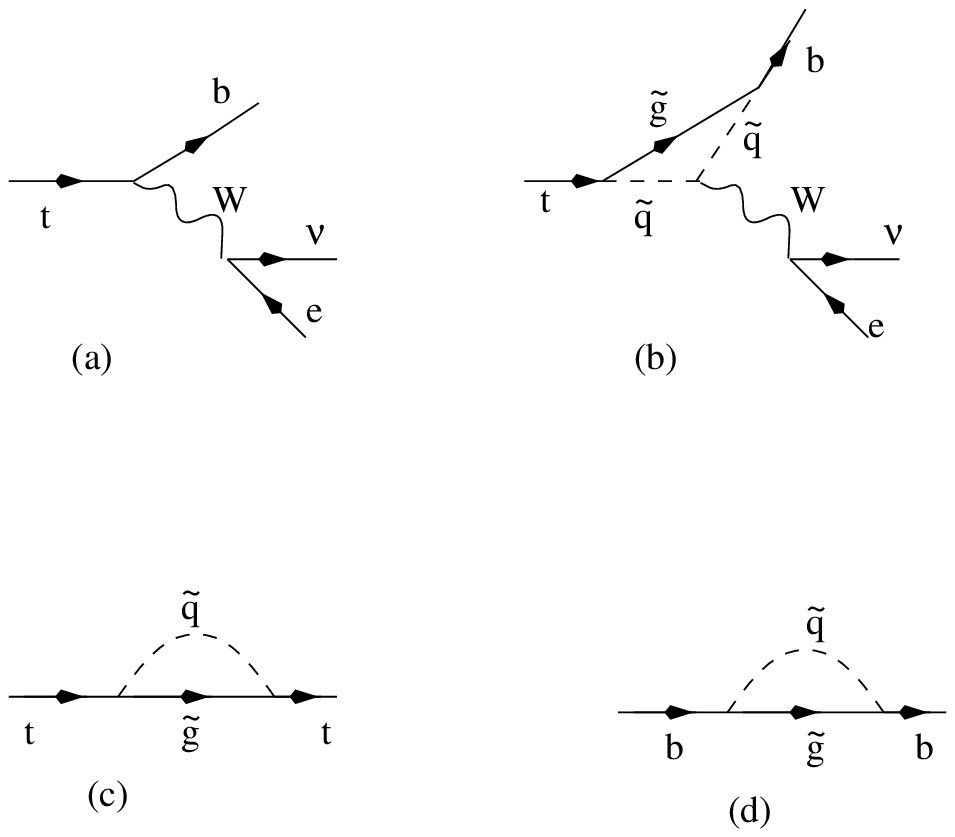}}
\caption[]{The tree and one-loop Feynman diagrams for the processes $t
\rightarrow b \bar l v_l$ within the SUSY QCD.}
\end{figure}

\begin{figure}
\epsfxsize=15 cm
\centerline{\epsffile{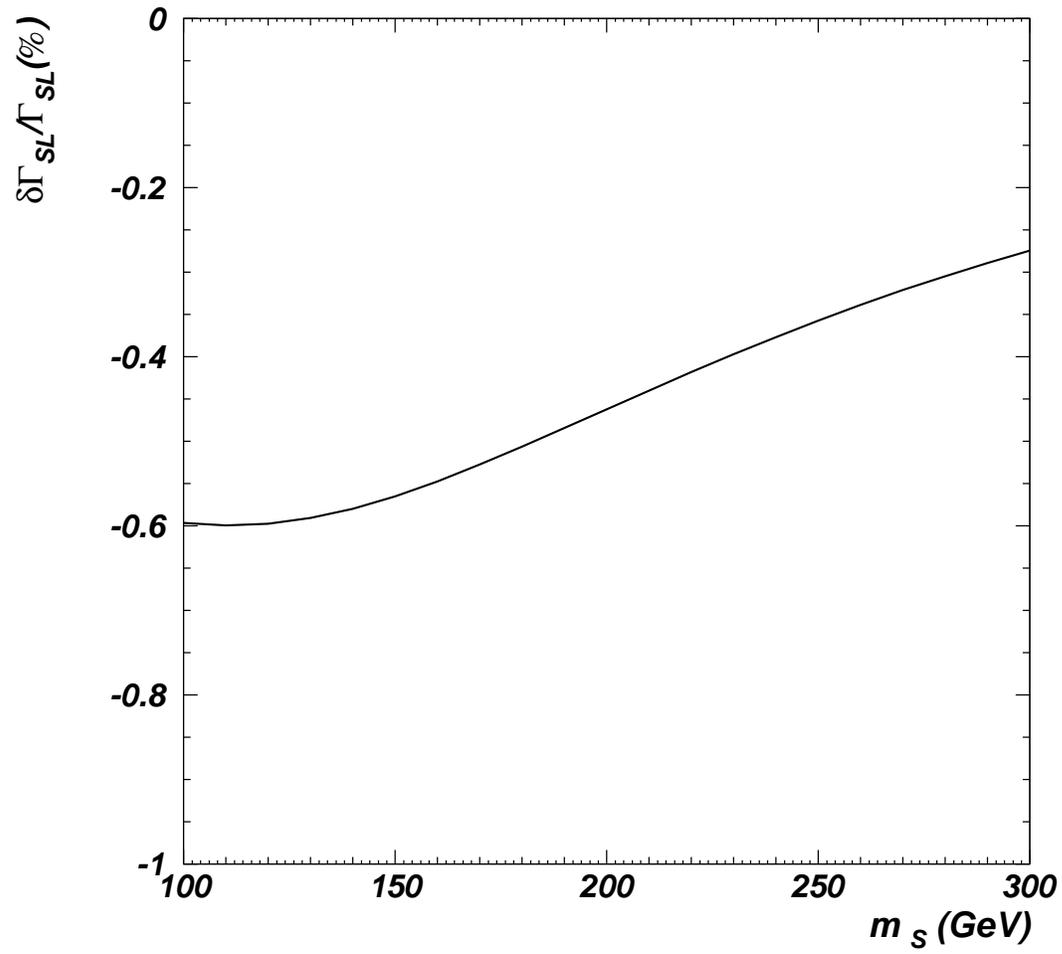}}
\caption[]{The relative correction as a function of $m_S$, while taking
$m_{\tilde{g}}=200$ GeV and $\tan\beta=2$.}
\end{figure}

\begin{figure}
\epsfxsize=15 cm
\centerline{\epsffile{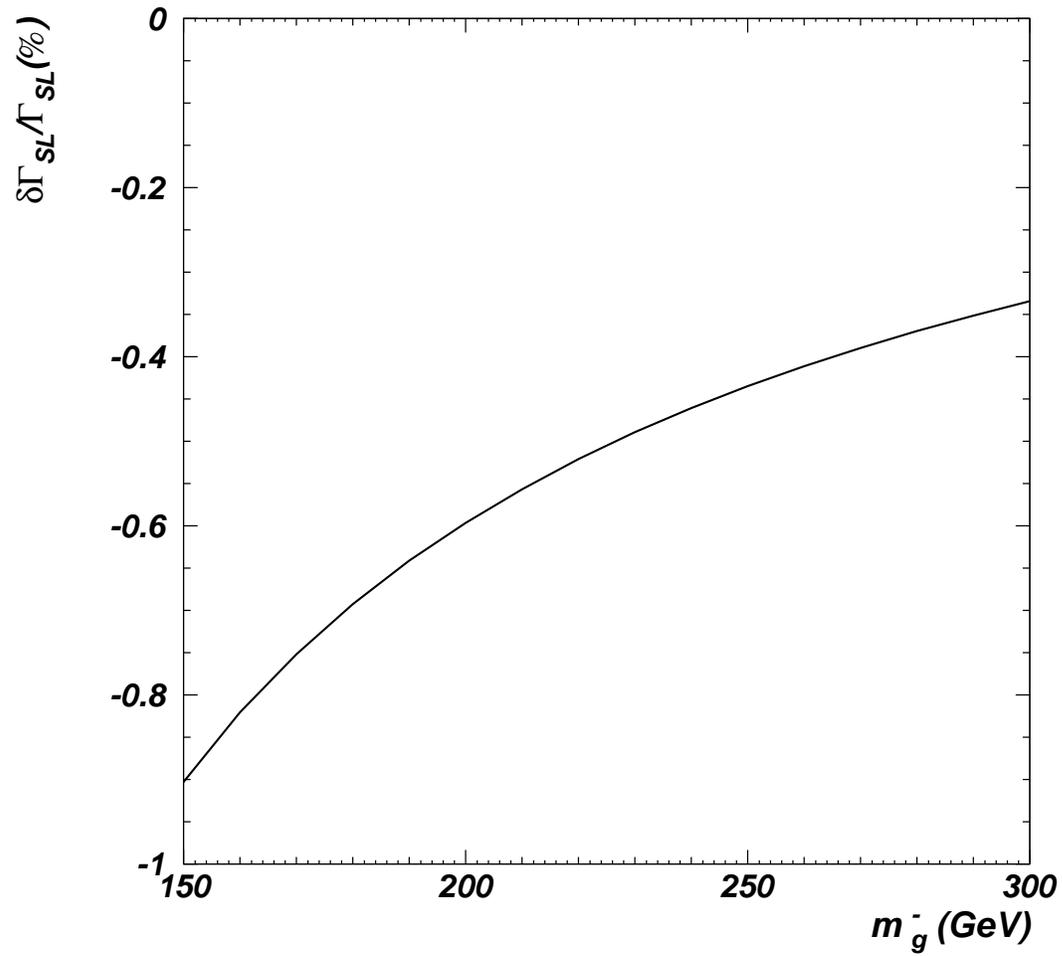}}
\caption[]{The relative correction as a function of $m_{\tilde{g}}$, while
taking $m_{S}=100$ GeV and $\tan\beta=2$.}
\end{figure}

\begin{figure}
\epsfxsize=15 cm
\centerline{\epsffile{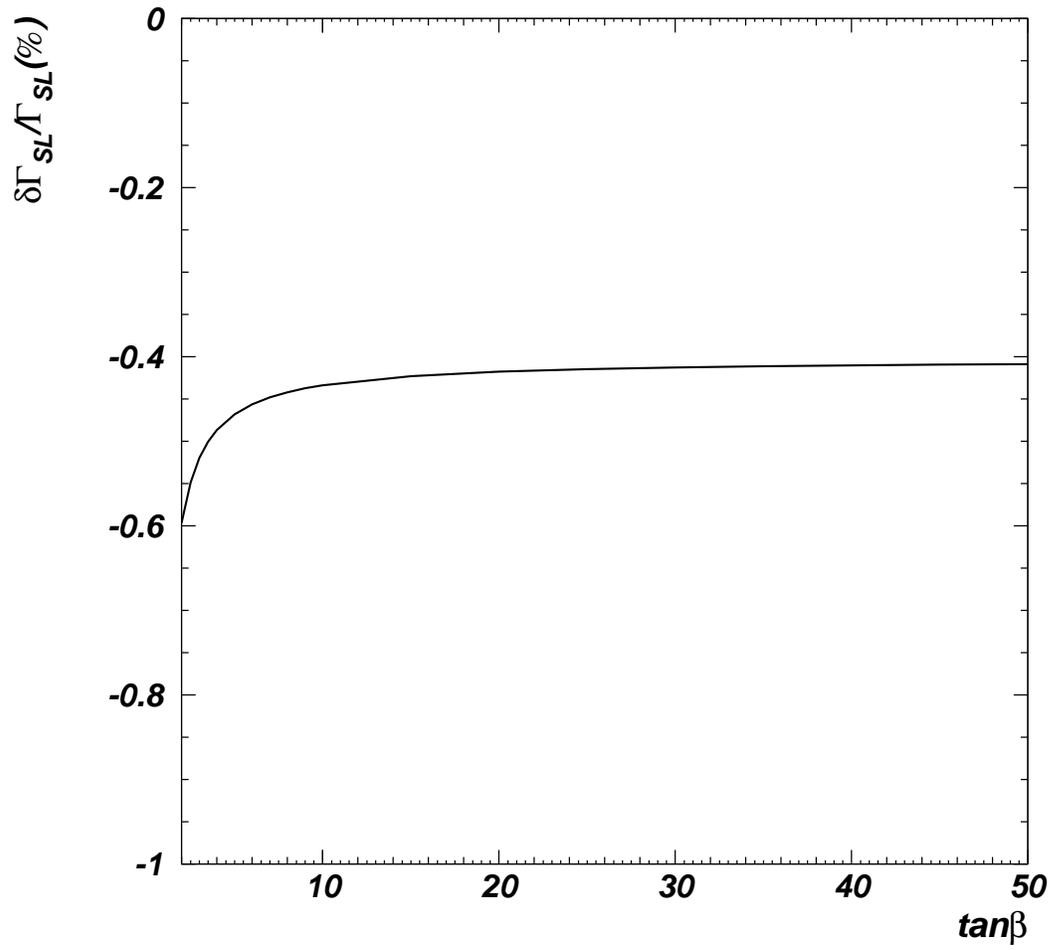}}
\caption[]{The relative correction as a function of $\tan\beta$, while
taking $m_{S}=100$ GeV and $m_{\tilde{g}}=200$ GeV.}
\end{figure}

\end{document}